\documentclass[journal]{IEEEtran}
\ifCLASSINFOpdf
\else
\fi
\usepackage{graphicx}
\usepackage{multicol,multirow}
\usepackage{amsmath,amssymb,amsfonts, bm}
\usepackage{mathrsfs}
\usepackage{amsthm}
\usepackage{rotating}
\usepackage[numbers]{natbib}
\usepackage{ifpdf}
\usepackage[T1]{fontenc}
\usepackage{newtxtext}
\usepackage{newtxmath}
\usepackage{textcomp}
\usepackage{xcolor}
\usepackage{lipsum}
\usepackage{url}
\usepackage{subfiles}
\usepackage{enumitem}
\usepackage{subcaption}
\usepackage{nccmath, mathtools}
\usepackage{dirtytalk}

\usepackage{romannum}
\usepackage{nicematrix}
\usepackage{stmaryrd}
\usepackage{caption} 
\usepackage{nccmath, mathtools}

\newtheorem{theorem}{Theorem}

\newtheorem{definition}{Definition}
\newtheorem{prop}{Proposition}

\newcommand{\bp}{\mathbf{p}}
\newcommand{\br}{\mathbf{r}}
\newcommand{\f}{\mathbf{f}}

\newcommand{\bz}{\mathbf{z}}

\newcommand{\bw}{\mathbf{w}}
\newcommand{\bx}{\mathbf{x}}

\newcommand{\bu}{\mathbf{u}}
\newcommand{\bd}{\mathbf{d}}

\newcommand{\bB}{\mathbf{B}}
\newcommand{\bF}{\mathbf{F}}

\newcommand{\bmu}{\bm{\mu}}

\newcommand{\bxi}{\bm{\xi}}
\newcommand{\btheta}{\bm{\theta}}
\newcommand{\bomega}{\bm{\omega}}
\newcommand{\balpha}{\bm{\alpha}}
\newcommand{\bbeta}{\bm{\beta}}
\newcommand{\bgamma}{\bm{\gamma}}

\newcommand{\bdelta}{\bm{\delta}}
\newcommand{\bpsi}{\boldsymbol{\psi}}

\newcommand{\rld}{\text{rld}}
\newcommand{\res}{\text{res}}

\newcommand{\R}{\mathbb{R}}

\newcommand{\bP}{\mathcal{P}}

\newcommand{\bQ}{\mathcal{Q}}

\DeclareMathOperator*{\argmax}{arg\,max}

\begin{document}
%
\title{An Efficient Learning-Based Solver for Two-Stage DC Optimal Power Flow with Feasibility Guarantees}
%
%
%

\author{Ling~Zhang,
        Daniel~Tabas,
        and~Baosen~Zhang
\thanks{L. Zhang (zhangling@microsoft.com) is with Microsoft Research Asia; D. Tabas (daniel.tabas2014@gmail.com) is with Base Power; and B. Zhang (zhangbao@uw.edu) is with  was with  Department of Electrical and Computer Engineering, University of Washington, Seattle. The work was done when L. Zhang and D. Tabas was at the University of Washington. The authors were partially supported by NSF award ECCS-2023531}.%

}

%
%

\markboth{CSEE JOURNAL OF POWER AND ENERGY SYSTEMS,~Vol.~x, No.~x, September~2024}%
{Shell \MakeLowercase{\textit{et al.}}: Bare Demo of IEEEtran.cls for IEEE Journals}
%



\maketitle

\begin{abstract}
In this paper, we consider the scenario-based two-stage stochastic DC optimal power flow (OPF) problem for optimal and reliable dispatch when the load is facing uncertainty. Although this problem is a linear program, it remains computationally challenging to solve due to the large number of scenarios needed to accurately represent the uncertainties. To mitigate the computational issues, many techniques have been proposed to approximate the second-stage decisions so they can be dealt more efficiently. The challenge of finding good policies to approximate the second-stage decisions is that these solutions need to be feasible, which has been difficult to achieve with existing policies. 

To address these challenges, this paper proposes a learning method to solve the two-stage problem in a more efficient and optimal way. A technique called the gauge map is incorporated into the learning architecture design to guarantee the learned solutions' feasibility to the network constraints. Namely, we can design policies that are feed forward functions and only output feasible solutions. Simulation results on standard IEEE systems show that, compared to iterative solvers and the widely used affine policy, our proposed method not only learns solutions of good quality but also accelerates the computation by orders of magnitude.
\end{abstract}


%
\IEEEpeerreviewmaketitle

\section[Introduction]{Introduction}
The optimal power flow (OPF) problem is one of the fundamental tools in the operation and planning of power systems~\cite{dommel1968optimal,baldick2006applied,Glover17}.
It determines the minimum-cost generator outputs that meet the  system demand and satisfy the power flow equations and operational limits on generators, line flows and other devices.
Traditionally, the OPF is formulated as a deterministic optimization problem, where a solution is computed for some nominal and fixed demand. However, with significant penetration of renewable energy into the power grid as well as demand response programs, the fluctuation in the demand should be explicitly taken into account~\cite{Bienstock2014CCOPF}. 

To take uncertainties in the net-load into account,\footnote{In this paper, we use the term net-load to capture both renewable generation in the system~\cite{varaiya2011} and the load.}  stochastic programming methods are a type of common tools used to reformulate the OPF as a multi-stage problem~\cite{varaiya2011,rajagopal2011,zhang2014rld}. In these problems, decisions are made sequentially at each stage, based on the forecast of the net-load and the fact that additional adjustments can be made in future stages when the uncertainties are better known. 

In this paper, we consider a two-stage stochastic program based on the DC optimal power flow (DCOPF) model. The DC power flow model linearizes the power flow equations and is the workhorse in power industries~\cite{stott2009dc}. The two-stage DCOPF problem is also becoming increasingly popular as a canonical problem that incorporates the impact of uncertainties arising from renewable resources~\cite{sjodin2012risk,roald2016corrective}. In more general terms, the two-stage DCOPF problem falls under the category of \textit{two-stage stochastic linear programs with (fixed) recourse} (2S-SLPR)~\cite{birge1997SPsurvey}.





Like other 2S-SLPR problems, the second stage of the two-stage DCOPF involves an expectation of the uncertain parameters ( i.e., the randomness in the net-load) over some probability distribution. In practice, the probability distributions are rarely known and difficult to work with analytically. Therefore, several different approaches have been used to approximate 2S-SLPR problems. Among these, the most popular is the sample average approximation (SAA)~\cite{shapiro2000,shapiro2009lecturesSP, birge2011introSP}. 

The SAA is a basic Monte Carlo simulation method, which represents the random parameter using a finite set of realizations (scenarios), yielding a (possibly large) deterministic two-stage linear programming problem. Though the SAA approach is easy to implement, directly using it to solve two-stage DCOPF may result in computational challenges. A reason for this is that the SAA method tends to require a large sample set in order to generate a good-quality solution~\cite{linderoth2006saabehavior,homemdemello2014surveyMonteCarlo,kim2015saa}, rendering the SAA formulation for two-stage DCOPF into a very large-scale linear program. In some sense, the challenge has moved from generating many high-quality samples from a probabilistic forecast to being able to solve an optimization problem using these samples~\cite{Chen18,pan2019deepopfa,Fioretto:IJCAI22c}.  
Secondly, as decisions in power system operations are made in a more online (or corrective) manner~\cite{roald2016corrective,capitanescu2011state}, OPF problems need to be solved repeatedly in real time. Even though solving single linear programs is easy, solving two-stage DCOPF problems are not~\cite{zhang2020cvxsolver,pan2019deepopfa}. 

A common approach to reduce the computational burden in solving two-stage DCOPFs is to model the second-stage decisions using an affine policy. More specifically, the second-stage (or the recourse) dispatch decision is restricted to be an affine function of the realized net-load and the first-stage decisions~\cite{Kuhn2011PrimalAD,Vrakopoulou2013ProbabilisticGF,roald2023}. 
Once the affine policy is determined, the decision-making in the real time is just simple function evaluations.
This method has been observed to provide good performance when the net-load variations are small or are restricted to a few possible instances~\cite{zhang2017drccopf,Vrakopoulou2017reserve,kannan2019}. However, if the variations are large or include many possible values, the affine policy method tends to not perform well. In fact, it may produce decisions that do not even satisfy the constraints in the two-stage optimization problem. 

In this paper, we overcome the challenge in policy design and solve two-stage DCOPF problems by presenting a neural network (NN)-based architecture that is computationally efficient and also guarantees the feasibility of learned solutions.  In particular, our architecture involves two neural networks, one each for the first and second stages. The first neural network learns the mapping from the load forecast to the first-stage decisions. The second neural network approximates the cost-to-go given the net-load realization and the learned first-stage decisions. 
So, instead of using the affine policy, we offer an NN-based policy to solve the second-stage OPF problem. 

Recently, many machine-learning based algorithms have been proposed to accelerate the computational speed for OPF, please see~\cite{pan2019deepopfa,deka2019learning,Duchesne20,Singh20,Donti21} and the references within. However, many of the existing algorithms are not suitable in a two-stage problem. To enhance computational speed, the proxy of the second stage cannot involve multiple steps (e.g., a neural network followed by a power flow solver). For feasibility, the solutions need to have feasibility guarantees. In this paper, we use an NN policy that is constructed using a technique called the \textit{gauge map}~\cite{Tabas2021ComputationallyES,TabasCDC,Li2023hardlinear}, which allows the output of the NN to be guaranteed to satisfy the DCOPF constraints. Since this policy also involves only function evaluations, it preserves the speed of affine policies. At the same time, a neural network is much more expressive than an affine function, and can provide much better approximations to the true solution. 


The main advantages of the proposed learning architecture are summarized below:
\begin{itemize}
\setlength{\itemindent}{1.5em}
    \item[1)] Since decision-making using the NNs only involves feedforward calculations, the proposed approach can solve problems at much faster speed (i.e., within milliseconds on average) compared to iterative solvers.
    \item[2)] By using the gauge map, the neural networks' outputs are guaranteed to be feasible solutions in the constraint set. As a result, all constraints in the problem are satisfied by construction, which cannot be done using affine policies or other existing learning approaches that uses multiple neural networks for multiple-stage problems~\cite{Duchesne20,Singh22}.  
    \item[3)] Since the gauge map guarantees the feasibility of solutions, the neural networks do not need to rely on the labeled data. Instead, the training dataset contains just the load scenario forecasts. As data sampling incurs minimal computational cost, this unsupervised training procedure can significantly improve the sample efficiency.
     \item[4)] We validate the effectiveness of the proposed approach by applying it to solving two-stage DCOPF problems on the 118-bus and 2000-bus systems. The simulation results demonstrate the ability of our approach to generate high-quality solutions orders of magnitude more quickly than commercial solvers.
\end{itemize}

The rest of this article is organized as follows:
In Section \ref{sec:setup}, we describe
the general setup of the two-stage DCOPF problem
and 
the two widely-used formulations of two-stage DCOPF.
Section \ref{sec:algorithm} presents the proposed learning approach to 
solve the two-stage DCOPF problem, including the overall architecture design, the training of it and the decision-making procedure.
Section \ref{sec:arch} illustrates how to incorporate the gauge map technique into the architecture design to ensure the feasibility of the neural networks' predictions. Section \ref{sec:results} provides the simulation results and Section \ref{sec:conclusion}  concludes the paper.

\section{Two-stage DCOPF}\label{sec:setup}
In this section, we provide more details about the formulation of two-stage DCOPF problems. 
Consider a power network with $N$ buses connected by $M$ transmission lines. 
Without loss of generality, we assume each bus $i$ has a generator as well as a load, and the load is uncertain. 
We denote the randomness in the system by $\bomega\in\R^{N}$, which is a random vector, and the net-load at each bus $i$ is a function of $\bomega$, denoted by $d_i(\bomega)$. Note that this notation allows us to capture the fact that the load depends non-trivially on the underlying randomness. The algorithm developed in this paper is compatible with any scenario-based forecasting algorithms. 


In the first stage of a problem, the exact value of $d_i(\bomega)$ is not known. Rather, we assume a forecast is available. Specifically, we adopt a scenario-based probabilistic load forecasting framework in this paper and assume 
a set of samples (scenarios) that are representative of $\bomega$ are available~\cite{Hong2016Tutorial, Khoshrou2019ShorttermSP, Gu2018GANbasedRL, Hong2018PLF, zhang2019flow}. It is useful to assume that a nominal load--for example, the mean of $d_i(\bomega)$--is known in the first stage. We denote this nominal load by $\widebar{d}_i$, and based on the scenario forecasts and $\widebar{d}_i$, the system operator (SO) chooses a first-stage generation dispatching decision, denoted by  $p_i^{0}$. Then once the actual demand $d_i(\bomega)$ is realized, a second-stage (recourse) decision $p_i^R$ is determined to balance the power network.


For concreteness, we specifically consider two widely used formulations of the two-stage DCOPF problem, risk-limiting dispatch (RLD)~\cite{varaiya2011} and reserve scheduling~\cite{Vrakopoulou2017reserve}. Both are two-stage stochastic linear programs with recourse, and both highlight the structure and difficulty of two-stage problems. 

\subsection{Risk-Limiting Dispatch}
The RLD problem seeks to find  a first-stage dispatching decision $p_i^{0}$ at each bus $i$ that minimizes expected total cost in two stages. The second-stage decisions, $p_i^{R}$, are made after the net-load is observed.



We assume that the cost of dispatching generation at bus $i$ is $\alpha_i p_i^{0}$ in the first stage and $\beta_i[p_i^{R}]^{+}$ in the second stage, where $\alpha_i$ and $\beta_i$ are prices measured in dollars per MW ($\$$/MW) and the notation $[z]^{+}=\max\{z,0\}$ means that only power purchasing ($p_i^{R}>0$) incurs a second-stage cost and any excess power ($p_i^{R}<0$) can be disposed of for free~\cite{varaiya2011,garcia2019approximating,palma2013modelling}. The cost minimization problem is:
\begin{subequations} \label{origModel-s1}
\begin{align} 
J_{\rld}^{\star}(\widebar{\bd})\triangleq\min_{\bp^{0}} \quad & \balpha^T\bp^{0} + \mathbb{E}[Q( \bd(\bomega)-\bp^{0};\bbeta)|\widebar{\bd}]\label{1o}\\
\text{s.t.} \quad & \bp^{0} \geq \mathbf{0} \label{1a},
\end{align}
\end{subequations}
where the expectation is taken with respect to the probability distribution of $\bd(\bomega)$ conditioned on $\widebar{\bd}$, and
$Q(\bd(\bomega)-\bp^{0};\bbeta)$ is the second-stage cost or cost-to-go.
Given the first-stage decision $\bp^0=[p_1^{0},\cdots,p_N^{0}]^T$ and a particular realization of $\bd(\bomega)$, the second-stage cost is given by
\begin{subequations} \label{origModel-s2}
\begin{align} 
    Q(\bd(\bomega)-\bp^{0};\bbeta)\triangleq \min_{\bp^{R}, \btheta} \quad & \bbeta^T[\bp^{R}]^{+} \\
    \text{s.t.} \quad & \bB\btheta = \bp^{R} - (\bd(\bomega)-\bp^{0})\label{2a}\\
    & -\f^{\max} \leq \bF\btheta \leq \f^{\max}, \label{2b}\\
    & {{\bpsi}^{\min} \leq \mathbf{E}^T \btheta \leq {\bpsi}^{\max}}\label{2c},
\end{align}
\end{subequations}
where 
\eqref{2a} is the DC power flow constraints, \eqref{2b} is the line flow limit constraints {and \eqref{2c} is angle difference limit constraints}. 
Without loss of generality, we assume bus $1$ is the reference (slack) node and set its voltage angle to be zero. The notation $\btheta\in\R^{N-1}$ denotes the voltage angles at non-slack buses,
the matrix $\bB\in\R^{N\times(N-1)}$ maps $\btheta$ to nodal power injections, the matrix $\bF\in\R^{M\times(N-1)}$ maps $\btheta$ to the flows on all edges, and the matrix $\mathbf{E}\in\R^{(N-1)\times M}$ is the incidence matrix of the network graph.
See Appendix \ref{appendix:BF} for details on constructing $\bB$, $\bF$ and $\mathbf{E}$.

Note that the second-stage problem \eqref{origModel-s2} can be seen as a deterministic DCOPF problem with the demand $\bd(\bomega)-\bp^{0}$. Since the recourse decision $\bp^{R}$ is not bounded, \eqref{origModel-s2} is feasible for 
any given demand input.



We approximate the expectation in \eqref{origModel-s1} using samples. 
Let $\{\bomega^{k}\}_{k=1}^{K}$ be a collection of 
samples of $\bomega$, and $\{\bd({\bomega}^{k})\}_{k=1}^{K}$ be the collection of load realizations. We determine the first-stage decision by solving the following scenario-based problem that is a deterministic linear program:
\begin{subequations}\label{saa-origModel}
\begin{align}
\widetilde{J}^{K}_{\rld}(\widebar{\bd})\triangleq
    & {\min_{
\substack{\bp^{0},\\ \{{\bp^{R}}({\bomega}^{k}),{\btheta}({\bomega}^{k})\}_{k=1}^{K}}
}~\balpha^T\bp^{0} + \frac{1}{K}\sum_{k=1}^{K} \bbeta^T[{\bp^{R}}({\bomega}^{k})]^{+}}   \label{3o} \\
\text{s.t.} \quad & \bp^{0} \geq \mathbf{0}\label{3a}\\
& \bB\btheta({\bomega}^{k}) = {\bp^{R}}({\bomega}^{k}) - ({\bd({\bomega}^{k})}-\bp^{0}), \; \forall k \label{3b}\\
& -\f^{\max} \leq \bF\btheta({\bomega}^{k}) \leq \f^{\max}, \forall k \label{3c}\\
& {{\bpsi}^{\min} \leq \mathbf{E}^T \btheta({\bomega}^{k}) \leq {\bpsi}^{\max}}, \forall k\label{3d}
\end{align}
\end{subequations}
where
the second-stage decisions $\{{\bp^{R}}({\bomega}^{k}),{\btheta}({\bomega}^{k})\}_{k=1}^{K}$ are functions of $\bomega$ and the constraints \eqref{3b}-\eqref{3d} related to the second-stage decisions need to be satisfied for every load realization $\bd({\bomega}^{k})$.

\subsection{Two-stage DCOPF with Reserve}\label{sec:II-B}
Sometimes the second-stage recourse decision $\bp^{R}$ are constrained by various factors such as generator capacities or  market rules. This is captured by a two-stage DCOPF where reserve services are  provided to deal with the possible mismatch between the prescheduled generation and the realized load~\cite{roald2016corrective,Vrakopoulou2017reserve}.

Specifically, we consider the spinning reserve service in this paper.
In the first stage, in addition to choosing an initial dispatching decision $p_i^0$ at each bus $i$, the SO also needs to decide the up and down reserve capacities, denoted by $\hat{r}_i$ and $\check{r}_i$, respectively. In this way, the first-stage cost at each bus $i$  includes both the cost of 
dispatching $p_i^{0}$, i.e.,  $\alpha_i p_i^{0}$, and of providing reserve services that is given by $\mu_i(\hat{r}_i+\check{r}_i)$, where $\alpha_i$ and $\mu_i$ are prices measured in $\$$/MW. 

The second-stage recourse decision $p^{R}_i$ at each bus $i$ is constrained by the reserve capacities, $\hat{r}_i$ and $\check{r}_i$. To quantify the 
amount by which the reserve capacities decided in the first stage might be exceeded, we define the cost of dispatching $p_i^R$ at each bus $i$ as a piecewise-affine function given by
$ \gamma_i^{\res}\Big([p_i^R-\hat{r}_i]^{+} - [p_i^R+\check{r}_i]^{-}\Big)$, where $\gamma_i^{\res}$ is penalty cost in $\$$/MW and $[z]^{-}=\min\{z,0\}$. 
This cost function means that there would be no cost for second-stage dispatching within the reserve amounts that are allocated in the first stage.

The two-stage DCOPF with reserve scheduling can be formulated as the following stochastic program:
\begin{subequations} \label{resModel-s1}
\begin{align} 
& J_{\res}^{\star}(\widebar{\bd})\triangleq \nonumber\\
\min_{\substack{\bp^{0},\\ \hat{\br},\check{\br}}}\quad & \balpha^T\bp^{0} +\bmu^T(\hat{\br}+\check{\br}) + 
\mathbb{E}[Q( \bd(\bomega)-\bp^{0};\hat{\br},\check{\br}, \bgamma^{\res})|\widebar{\bd}]\\
\textrm{s.t.} \quad &  \mathbf{0}\leq \bp^{0} \leq {\bp}^{\max}\label{4a}\\
&\bp^{0} + \hat{\br} \leq  \bp^{\max} \label{4b}\\
&\bp^{0} - \check{\br} \geq \mathbf{0} \label{4c}\\
& \hat{\br},~\check{\br}\geq \mathbf{0} \label{4d},
\end{align}
\end{subequations}
where 
\eqref{4b}-\eqref{4d} constrain the up and down reserve at each bus $i$ to be positive and no larger than the available capacities 
around $p_i^0$.
Given the first-stage decisions $(\bp^{0},\hat{\br},\check{\br})$ and a particular realization of $\bd(\bomega)$, the second-stage cost is given by 
\begin{subequations} \label{resModel-s2}
\begin{align} 
& Q(\bd(\bomega)-\bp^{0};\hat{\br},\check{\br}, \bgamma^{\res})\triangleq\nonumber\\
\min_{\bp^{R}, \btheta} \quad & {\bgamma^{\res}}^{T}\Big( [\bp^R-\hat{\br}]^{+}-
[\bp^R+\check{\br}]^{-} \Big) \\
\textrm{s.t.} \quad & \bB\btheta = \bp^{R} - (\bd(\bomega)-\bp^{0})\label{5a}\\
& -\f^{\max} \leq \bF\btheta \leq \f^{\max}\label{5b}\\
& {{\bpsi}^{\min} \leq \mathbf{E}^T \btheta \leq {\bpsi}^{\max}}\label{5c},
\end{align}
\end{subequations}
which can also be seen as a deterministic DCOPF problem with demands $\bd(\bomega)-\bp^{0}$ and the cost being the penalty imposed on the generation amount if it exceeds the reserve capacities.
This \say{penalizing deviations} technique is commonly employed by stochastic programmers to promote the feasibility of second-stage problems for any given first-stage decisions~\cite{higle2005}.

The SAA method solves the following scenario-based problem associated with \eqref{resModel-s1}:
\begin{subequations} \label{saa-resModel}
\begin{align} 
\widetilde{J}^{K}_{\res}(\widebar{\bd})\triangleq & \min_{
\substack{\bp^{0},\hat{\br},\check{\br},\\ \{{\bp^{R}}({\bomega}^{k}),{\btheta}({\bomega}^{k})\}_{k=1}^{K}}
} \quad \balpha^T\bp^{0} + \bmu^T(\hat{\br}+\check{\br})~+\nonumber\\
\quad & \frac{1}{K}\sum_{k=1}^{K} \Bigg(
{\bgamma^{\res}}^{T} \Big( [\bp^R({\bomega}^{k})-\hat{\br}]^{+}-
[\bp^R({\bomega}^{k})+\check{\br}]^{-} \Big)
\Bigg)\\
\textrm{s.t.}\quad & \mathbf{0}\leq \bp^{0} \leq {\bp}^{\max}\label{6a}\\
&\bp^{0} + \hat{\br} \leq  \bp^{\max} \label{6b}\\
&\bp^{0} -\check{\br} \geq \mathbf{0} \label{6c}\\
& \hat{\br},~\check{\br}\geq \mathbf{0} \label{6d}\\
& \bB\btheta({\bomega}^{k}) = {\bp^{R}}({\bomega}^{k}) - ({\bd({\bomega}^{k})}-\bp^{0}), \forall k\label{6e}\\
& -\f^{\max} \leq \bF\btheta({\bomega}^{k}) \leq \f^{\max}, \forall k \label{6f}\\
& {{\bpsi}^{\min} \leq \mathbf{E}^T \btheta({\bomega}^{k}) \leq {\bpsi}^{\max}}, \forall k.\label{6g}
\end{align}
\end{subequations}

\subsection{Computational Challenges}
To have a sample set of load realizations that is representative enough of the true distribution of the random net-load, a large number of realizations are required for even a moderately sized system~\cite{ShapRusz03}.
Therefore, although \eqref{saa-origModel} and \eqref{saa-resModel} are linear programs, they are often large-scale problems. In addition, since both the first and second-stage decisions depend on the mean of the scenario forecasts, $\widebar{\bd}$, every time the set of scenarios changes, we need to re-solve \eqref{saa-origModel} and \eqref{saa-resModel}.
Even if a single instance can be solved efficiently using commercial solvers such as CVXPY~\cite{diamond2016cvxpy,agrawal2018rewriting} and GLPK~\cite{Oki2012GLPKL}, repeatedly solving large-scale linear programs can still impose considerable computational burdens.

The scale of the problems can grow quickly as the size of the system and the number of scenarios grow. Therefore, an affine policy is often used to  approximate \eqref{origModel-s2} and \eqref{resModel-s2}. However, finding a good policy that satisfies the constraints \eqref{3b},\eqref{3c}, \eqref{3d}, \eqref{6e}, \eqref{6f}, and \eqref{6g} can be difficult.  In the next section, we present an NN-based learning architecture to enable more efficient computation.

\section{Proposed Learning Algorithm }\label{sec:algorithm}
In this section, we present the learning algorithm to solve the scenario-based problems in \eqref{saa-origModel} and \eqref{saa-resModel}.
To start with, we rewrite the two-stage problem in a more compact way as follows
\begin{subequations} \label{generic-s1}
\begin{align} 
\widetilde{J}^{K}(\widebar{\bd})\triangleq \quad & \min_{
\bx} \quad \widetilde{c}(\bx) + 
\widetilde{Q}^{K}(\bx; \{\bomega^{k}\}_{k=1}^{K},\widetilde{\bbeta}) \\
\textrm{s.t.} \quad & \bx\in\mathcal{X}\label{7a}
\end{align}
\end{subequations}
where $\bx$ denotes the first-stage decisions, which is $\bp^0$ for \eqref{saa-origModel} and  $(\bp^{0},\widehat{\br},\check{\br})$ for \eqref{saa-resModel}, and the set $\mathcal{X}$ collects all constraints that $\bx$ has to satisfy, i.e., 
\eqref{3a} or \eqref{6a}-\eqref{6d}. The notation
$\widetilde{c}(\cdot)$ is the generic representation of the first-stage cost
and 
$\widetilde{Q}^{K}(\bx; \{\bomega^{k}\}_{k=1}^{K},\widetilde{\bbeta})$
is the estimated second-stage cost based on a collection of scenarios 
$\{\bd({\bomega}^{k})\}_{k=1}^{K}$.

Here we use a simple decomposition technique such that \eqref{generic-s1} becomes much easier to work with.
To be specific, if 
the first-stage decision $\bx$ is taken as given, then the second-stage cost $\widetilde{Q}^{K}(\bx; \{\bomega^{k}\}_{k=1}^{K},\widetilde{\bbeta})$ is separable:
\begin{equation}
    \widetilde{Q}^{K}(\bx; \{\bomega^{k}\}_{k=1}^{K},\widetilde{\bbeta}) = \frac{1}{K}\sum_{k=1}^{K} 
\widetilde{Q}^{k}(\bdelta_d(\bx; {\bomega}^{k});\bx,\widetilde{\bbeta})\nonumber
\end{equation}
where we use the notation $\bdelta_d(\bx;{\bomega}^{k})\triangleq\bd({\bomega}^{k})-\bp^{0}$ to represent the demands that are not balanced by the first stage when the load realization is actually $\bd({\bomega}^{k})$, and 
$\widetilde{Q}^{k}(\bdelta_d(\bx; {\bomega}^{k});\bx,\widetilde{\bbeta})$  
is the optimal value of each scenario problem for a particular load realization $\bd({\bomega}^{k})$.
Each of these scenario problems can be seen as a deterministic DCOPF problem with demands 
$\bdelta_d(\bx;{\bomega}^{k})$ and an objective function $\widetilde{q}(\cdot; \bx,\widetilde{\bbeta})$ that {takes $\bx$ as given and $\widetilde{\bbeta}$ as parameters}. The deterministic DCOPF problem can  be written in the following generic form:
\begin{subequations} \label{generic-s2-k}
\begin{align} 
\widetilde{Q}(\bdelta_d(\bx;\bomega);\bx,\widetilde{\bbeta})\triangleq \min_{\bp^{R}, \btheta} \quad & \widetilde{q}(\bp^R; \bx,\widetilde{\bbeta}) \\
\textrm{s.t.} \quad & \bB\btheta = \bp^{R} - \bdelta_d(\bx;\bomega)\label{8a}\\
& -\f^{\max} \leq \bF\btheta \leq 
\f^{\max}\label{8b}\\
& {{\bpsi}^{\min} \leq \mathbf{E}^T \btheta \leq {\bpsi}^{\max}}.\label{8c}
\end{align}
\end{subequations}

\captionsetup[figure]{font=small,skip=2pt}
\begin{figure}[ht!]
\centering
\includegraphics[scale=0.32]{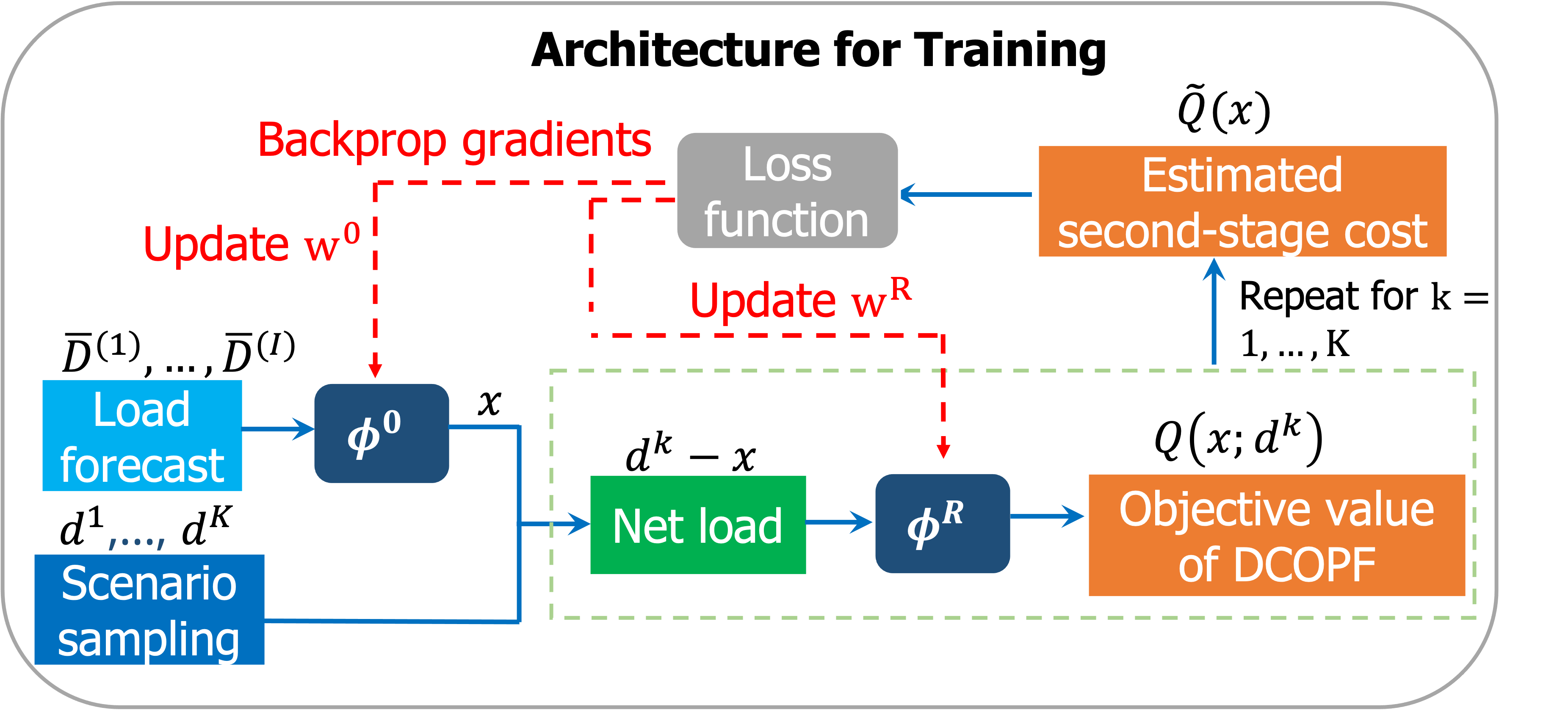}
\caption{The architecture used for training in the proposed algorithm. When making decisions in real time, we only need the network $\phi^{0}$ to predict the first-stage decisions from the given load forecast. }
\label{fig:architecture_train}
\end{figure}

In a similar fashion, by exploiting the decomposable structure of \eqref{generic-s1}, the proposed learning algorithm consists of two subnetworks, denoted by $\phi^{0}$ and $\phi^{R}$, respectively.
The first subnetwork $\phi^{0}$ learns the mapping from $\widebar{\bd}$ to $\bx$, i.e., $\phi^{0}:\R^{N}_{+}\longmapsto\mathcal{X}$, {while the second one learns the mapping from $\bdelta_d(\bx;\bomega)$ to
$\widetilde{Q}(\bdelta_d;\bx,\widetilde{\bbeta})$ considering $\bx$ as given, i.e.,  
$\phi^{R}: \Delta_d(\bx,\bomega) | \mathcal{X} \longmapsto [0,+\infty]$, } where $\Delta_d(\bx,\bomega)$ is the set of all possible mismatches, i.e., $\Delta_d(\bx,\bomega)=\{\bd(\bomega)-\bp^{0}|(\bx,\bomega)\in\mathcal{X} \times \Omega\}$, and $\Omega$ is the sample space of $\bomega$. 
{Note that the second network $\phi^{R}$ takes two inputs. The first input is the unbalanced demand $\bdelta_d(\bx;{\bomega}^{k})$ when the load realization is $\mathbf{d}({\bomega}^{k})$. The second input is the first stage decision $\bx$, which is treated as a given value. For the sake of analysis, we introduce an augmented vector $\tilde{\bdelta}_d$ that includes both inputs, specifically: $\tilde{\bdelta}_d = (\bdelta_d(\bx;{\bomega}^{k}), \mathbf{x})$.}



The two subnetworks can be implemented using neural networks.
Once trained, these neural networks can produce solutions much faster than existing solvers. 
However, a key question also arises:
how to make neural networks satisfy constraints, namely, how to ensure the output from $\phi^{0}$ lies within the feasibility set $\mathcal{X}$ and the constraints of the optimization problem in \eqref{generic-s2-k} are satisfied? Notably, we want to avoid steps such as projecting to the feasible set since these introduce additional optimization problems~\cite{agarwal2021theory}, which somewhat defeats the purpose of learning. This key question will be tackled in the next section, and for the rest of this section, we first treat the two subnetworks as black boxes to provide an overview of the proposed algorithm.

This algorithm includes a training process and a prediction process. The architecture used for training is shown in Fig.~\ref{fig:architecture_train}.
When the learning algorithm is used in practice, i.e., in the prediction phase, just the network $\phi^{0}$ is required to predict the first-stage decisions from a nominal load value.
The reason why we need a second network $\phi^{R}$ is that the two networks need to be trained together in order to obtain a network $\phi^{0}$ that can predict the solution to \eqref{generic-s1} accurately. 
We now describe how the two networks in 
Fig.~\ref{fig:architecture_train} are trained.

{Let $\bw^{0}$ and $\bw^{R}$ denote the respective parameters, i.e., the weights and biases, of neural networks $\phi^{0}$  and $\phi^{R}$.
The goal of training is to learn the optimal values for $\bw^{0}$ and $\bw^{R}$ that minimize a specifically designed loss function. Unlike standard machine learning practices, where the loss function is based on prediction errors using generated ground truth data, our model is trained in an unsupervised manner, which means the training dataset does not incorporate ground truth information. Instead, the parameters of $\phi^{0}$ and $\phi^{R}$ are updated to minimize the empirical mean of the two-stage problem's total cost.}

Concretely, given a batch of training data consisting of $I$ load forecasts, denoted by $\{\widebar{\bd}^{i}\}_{i=1}^{I}$, the training loss function for our model is formulated as follows: 
\begin{equation} 
\begin{split}
\min_{
\bw^{0},\bw^{R}}&L(\bw^{0},\bw^{R})\triangleq~
\frac{1}{I}\sum_{i=1}^{I} L^{i}(\bw^{0},\bw^{R}),
\end{split} \label{eq:train-loss}
\end{equation}
where
\begin{equation} 
\begin{split}
L^{i}(\bw^{0},\bw^{R})&\triangleq~
\widetilde{c}\Big(\phi^{0}(\widebar{\bd}^{i};\bw^{0})\Big) + 
\frac{1}{K}\sum_{k=1}^{K} {\phi^{R}\Big(\tilde{\bdelta}_d^{ik} ;\bw^{R}\Big)}.
\end{split} \nonumber
\end{equation}
We use the double superscript in $\tilde{\bdelta}_d^{ik} $ to represent that, for each instance of $\widebar{\bd}^{i}$, we need to sample an independent set of scenarios $\{\bomega^{ik}\}_{k=1}^{K}$.
{Note that the loss function in \eqref{eq:train-loss} represents exactly the average total cost of the two-stage DCOPF problem across the training dataset. In particular, the parameters of 
the first network $\phi^{0}$ are updated 
by jointly minimizing the first stage cost and the anticipated cost-to-go associated with its predictions. For the second network $\phi^{R}$, its parameters are updated to minimize the recourse cost across a collection of sampled load realizations. This is because, given a first stage decision $\bx$, the best decision that $\phi^{R}$ can make is actually the minimal cost one.}

{After formulating the loss function \eqref{eq:train-loss} during the forward pass, the gradients of this loss function with respect to $\bw^{0}$ and $\bw^{R}$ are calculated through the backward pass. Following that, the stochastic gradient descent (SGD) method is used to minimize the loss function. The formulas for updating $\bw^{0}$ and $\bw^{R}$ using SGD can be found in Appendix~\ref{appendix:sgd}.}


\captionsetup[table]{font=small,skip=2pt}
\begin{table}[t!]
\normalsize
\centering
\begin{tabular}{cl}
\hline
\multicolumn{2}{c}{\textbf{Proposed Learning Algorithm}}\\
\hline
\multicolumn{2}{l}{\textbf{Training Procedure}}\\
1:&\textbf{Inputs:}~Number of iterations $T$, a minibatch of \\
&training data, $\{\widebar{\bd}\}_{i=1}^{I}$, sample space $\Omega$ of $\bomega$ \\
2:&\textbf{Parameters:} $\bw^{0}$, $\bw^{R}$\\
3:&\textbf{for $t=1,\cdots,T$ do}\\
4:&\quad Randomly sample $\{\bomega^{ik}\}_{i=1:I,k=1:K}$ from $\Omega$.\\
5:&\quad Forward pass $\phi^{0}\Big(\{\widebar{\bd}\}_{i=1}^{I};\bw^{0}\Big)\longrightarrow \{\bx^{i}\}_{i=1}^{I}$\\
6:&\quad \textbf{for $k=1,\cdots,K$ do}\\
7:&\qquad\quad Calculate $\bdelta_d(\bx^{i},\bomega^{ik})$ for $i=1,\cdots,I$. \\
8:&\qquad\quad Forward pass $\phi^{R}\Big(\{\bdelta_d(\bx^{i},\bomega^{ik})\}_{i=1}^{I};\bw^{R}\Big)\longrightarrow$\\
&\qquad\quad $\Big\{\widetilde{Q}^{k}(\bdelta_d(\bx^{i};\bomega^{ik});\bx^{i},\tilde{\bbeta})\Big\}_{i=1}^{I}$\\
9:&\quad \textbf{end for}\\
10:&\quad Construct the loss function using \eqref{eq:train-loss}. \\
11:&\quad Randomly pick a data point $\widebar{\bd}^{i}$ and calculate the \\
&\quad stochastic gradients using \eqref{eq:sgd}.\\
12:&\quad Update $\bw^{0}$ and $\bw^{R}$ using \eqref{eq:sgd-update}.\\
13:&\quad Check stopping criterion.
\\
14:& \textbf{end for}\\
15:&\textbf{Outputs:} Trained networks $\phi^{0}$ and $\phi^{R}$\\
\hline
\multicolumn{2}{l}{\textbf{Decision Making Procedure}}\\
1:&\textbf{Inputs:}~Load forecast ${\widebar{\bd}}^{\text{new}}$, trained network $\phi^{0}$  \\
2:&Forward pass $\phi^{0}\Big({\widebar{\bd}}^\text{new};\bw^{0}\Big)\longrightarrow \bx^{\text{new}}$\\
3:&\textbf{Outputs:}~Predicted first-stage decision $\bx^{\text{new}}$\\ 
\hline
\end{tabular}
\caption{The proposed learning-based  algorithm to solve \eqref{generic-s1}.}
\label{tab:algo1}
\end{table}

Once parameters $\bw^{0}$ and $\bw^{R}$ reach a local minimum and the training process terminates, we can use the trained network $\phi^{0}$ that is parameterized by the learned $\bw^{0}$ to predict the first-stage decisions from load forecasts. 
We summarize our learning algorithm, including the training and the decision-making procedures, in Table~\ref{tab:algo1}. 

In the next section, we show the detailed architecture design of the two networks and answer the key question about how to make them satisfy the constraints in the optimization problems.

\section{Network Architecture Design}\label{sec:arch}
In this section, we show the network design of $\phi^{0}$ and $\phi^{R}$ that ensures the feasibility of the networks' outputs. Particularly, each network consists of a sequence of neural layers, which are convolutional or fully connected layers with an activation function 
applied after each layer, then followed by a series of transformations that 
map the output of the neural layers to a feasible solution. 

We first deal with first-stage constraints that must be satisfied by $\phi^{0}$. These constraints, as given in \eqref{3a} or \eqref{6a}-\eqref{6d}, describe axis-aligned rectangular regions and are easy to satisfy. We will then deal with the second-stage constraints that $\phi^R$ must satisfy. 
To be specific, $\phi^{R}$ is learning the optimal value of the second-stage DCOPF problem and must satisfy all the constraints in the optimization problem; otherwise, the estimated second-stage cost may have a large deviation from the true value and mislead the training of $\phi^{0}$. In turn, if the first-stage decisions are poorly made,
the resulting second-stage cost can be potentially very high when the uncertainties are realized.
In particular, the constraints in the second-stage problem describe a high-dimensional polyhedral set that is dependent on the input data; thus, guaranteeing feasibility requires some more nontrivial techniques. 



\subsection{Network Design of $\phi^{0}$}
The network $\phi^{0}$ in the RLD formulation must satisfy the non-negative orthant constraints in \eqref{3a}, which can be guaranteed by using a ReLU activation after the last neural layer, and no additional transformation is needed.\footnote{The ReLU activation function is $\max(x,0)$.} For the reserve scheduling problem, we can rewrite the constraints in \eqref{6a}-\eqref{6d} in a more compact way as $\bx\in[\widebar{\bx}, \underline{\bx}]$, where $\widebar{\bx}=[{{\bp}^{\max}}^{T},(\bp^{\max}-\bp^{0})^T,{\bp^{0}}^T]^T$ and $\underline{\bx}=[\mathbf{0}^T,\mathbf{0}^T,\mathbf{0}^T]^T$. 
In this way, the constraints in \eqref{6a}-\eqref{6d} can be treated as axis-aligned rectangular constraints.

\begin{figure}[t]
\centering
\includegraphics[width=8cm, height=4.5cm]{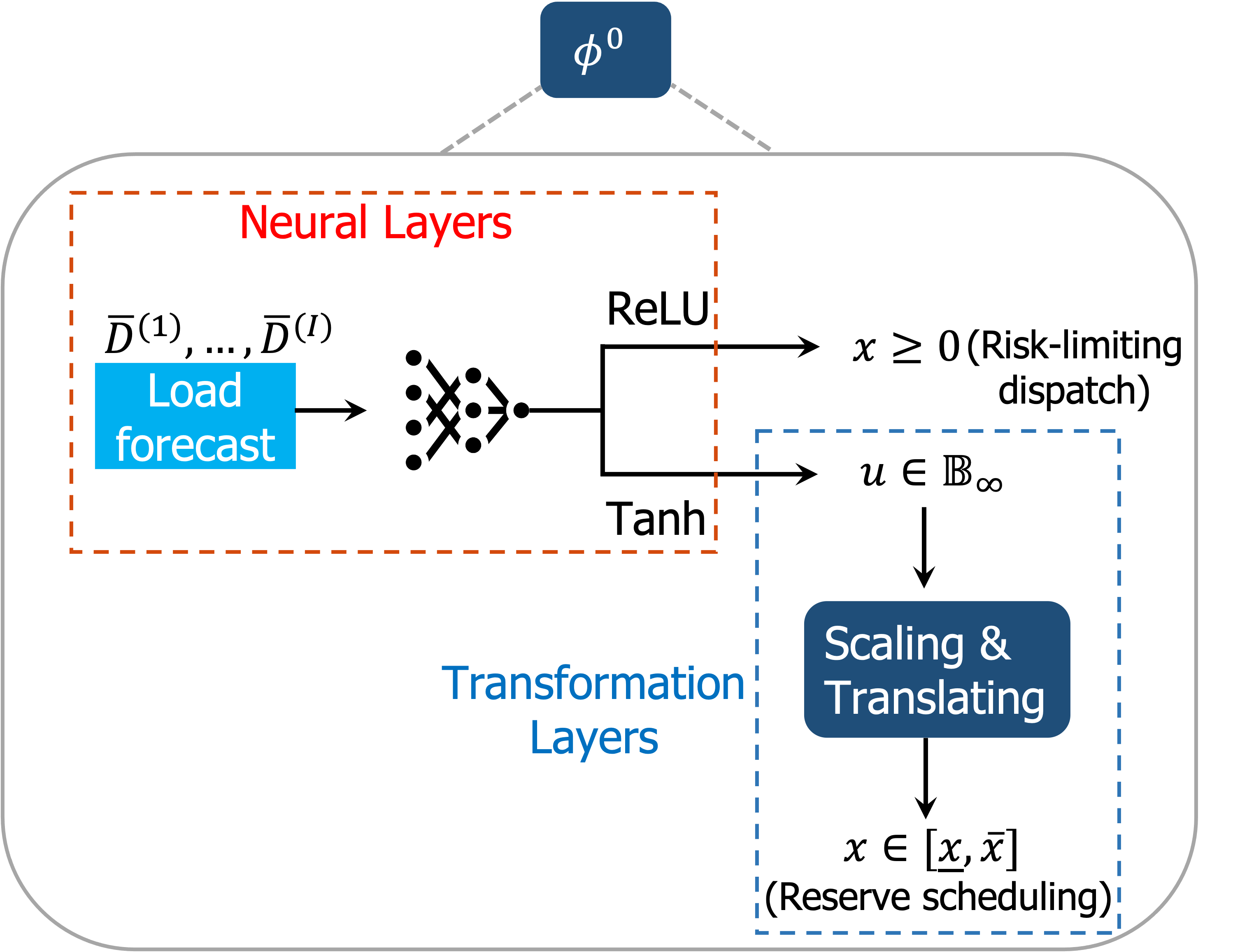}
\caption{In the RLD problem, non-negative orthant constraints can be enforced using ReLU activation in the last neural layer. For the reserve scheduling problem, the Tanh activation is used at the last neural layer and then the hypercubic output is passed through the transformation layers in
\eqref{phi0-mapping} to obtain a feasible solution.
\vspace{-0.5cm}}
\label{fig:archphi0}
\end{figure}

To enforce such axis-aligned rectangular constraints, we start by using a Tanh activation on the last neural layer and denote its output as $\bu$. Since the $\tanh$ function has a range between $-1$ and $1$, we have $\bu\in\mathbb{B}_{\infty}$, where $\mathbb{B}_{\infty}$ is the unit ball with $\ell_{\infty}$ norm given by $\mathbb{B}_{\infty}\triangleq\{\bz\in\R^{n}|-1\leq z_i\leq1,\forall i\}$. Next, we apply the following 
scaling and translating operations to transform $\bu$ to a feasible solution that satisfies  \eqref{6a}-\eqref{6d}:
\begin{equation}\label{phi0-mapping}
    {x}_i = \frac{1}{2}(u_i + 1) (\bar{x}_i-\underline{x}_i)+\underline{x}_i, \forall i.
\end{equation}
A diagram is provided in Fig.~\ref{fig:archphi0} to illustrate the separate network architectures of $\phi^{0}$ for \eqref{saa-origModel} and \eqref{saa-resModel}.
\captionsetup[figure]{font=small,skip=2pt}

\subsection{Network Design of $\phi^{R}$}
The network architecture design for $\phi^{R}$ is not as straightforward as for $\phi^{0}$ because the constraints 
in \eqref{8a}-\eqref{8c} can not be enforced by simply scaling and translating the neural layers' outputs.
{More importantly, these constraints involve two sets of variables, namely, $\bp^{R}$ and $\btheta$, which are interconnected by the equality constraints (the power flow equations) in \eqref{8a}. To this end, we can use \eqref{8a} to express the recourse variable $\bp^{R}$ as an affine function of $\btheta$. As a result, we can use $\phi^{R}$ only for predicting $\btheta$, and then reconstruct $\bp^{R}$ using the power flow equations. Therefore, the constraints that $\phi^{R}$ need to satisfy have been reduced to only $\eqref{8b}$ and $\eqref{8c}$, which define a polyhedral space of dimension $N-1$:}
\begin{equation}\label{def:Theta}
    \Theta = \left\{ \btheta|
   -\mathbf{\bar{f}} \leq \mathbf{F} \btheta \leq \mathbf{\bar{f}}, 
   {\bpsi}^{\min} \leq \mathbf{E}^T \btheta \leq {\bpsi}^{\max}
  \right\}.
\end{equation}
Next, we describe the architecture design of $\phi^{R}$ that transforms the output of neural layers to a point within $\Theta$.

Concretely,
we again use a Tanh activation function on the last neural layer and denote the output from it by $\bu$, which satisfies
$\bu\in\mathbb{B}_{\infty}$ as we have discussed. Then we utilize the  \textit{gauge map} technique~\cite{Tabas2021ComputationallyES} to fulfill the transformation. Particularly, the gauge map can establish the equivalence between two C-sets using the gauge functions associated with them.
{The definitions of C-sets and gauge functions are provided in \cite{Tabas2021ComputationallyES}, and we present them again here for the sake of clarity. We will also show that, by these definitions, both $\mathbb{B}_{\infty}$ and $\Theta$ are C-sets.}
\begin{definition}[C-set \cite{Blanchini07}]\label{def:Cset}
A C-set is a convex and compact subset of $\R^{n}$ including the origin as an interior point.
\end{definition}
By Definition \ref{def:Cset}, 
the unit hypercube $\mathbb{B}_{\infty}$ is a C-set. 
{To see that the polyhedral set $\Theta$ also satisfies Definition \ref{def:Cset}, we first note that the 
origin is already located within $\Theta$, as the origin satisfies the inequalities defining $\Theta$ in \eqref{def:Theta}.}
Regarding the compactness of $\Theta$, we provide the following theorem.
\begin{theorem}\label{thm:Cset}
    The polyhedral set $\Theta$ given by \eqref{def:Theta} is bounded.
\end{theorem}
The proof of Theorem \ref{thm:Cset} is given in Appendix \ref{proof:Cset}. Together, we can conclude that $\Theta$ is also a C-set. Before describing the gauge transformation between $\mathbb{B}_{\infty}$ and $\Theta$, we first introduce the concept of the gauge function associated with a C-set.

\begin{figure}[t]
\centering
\includegraphics[scale=0.45]{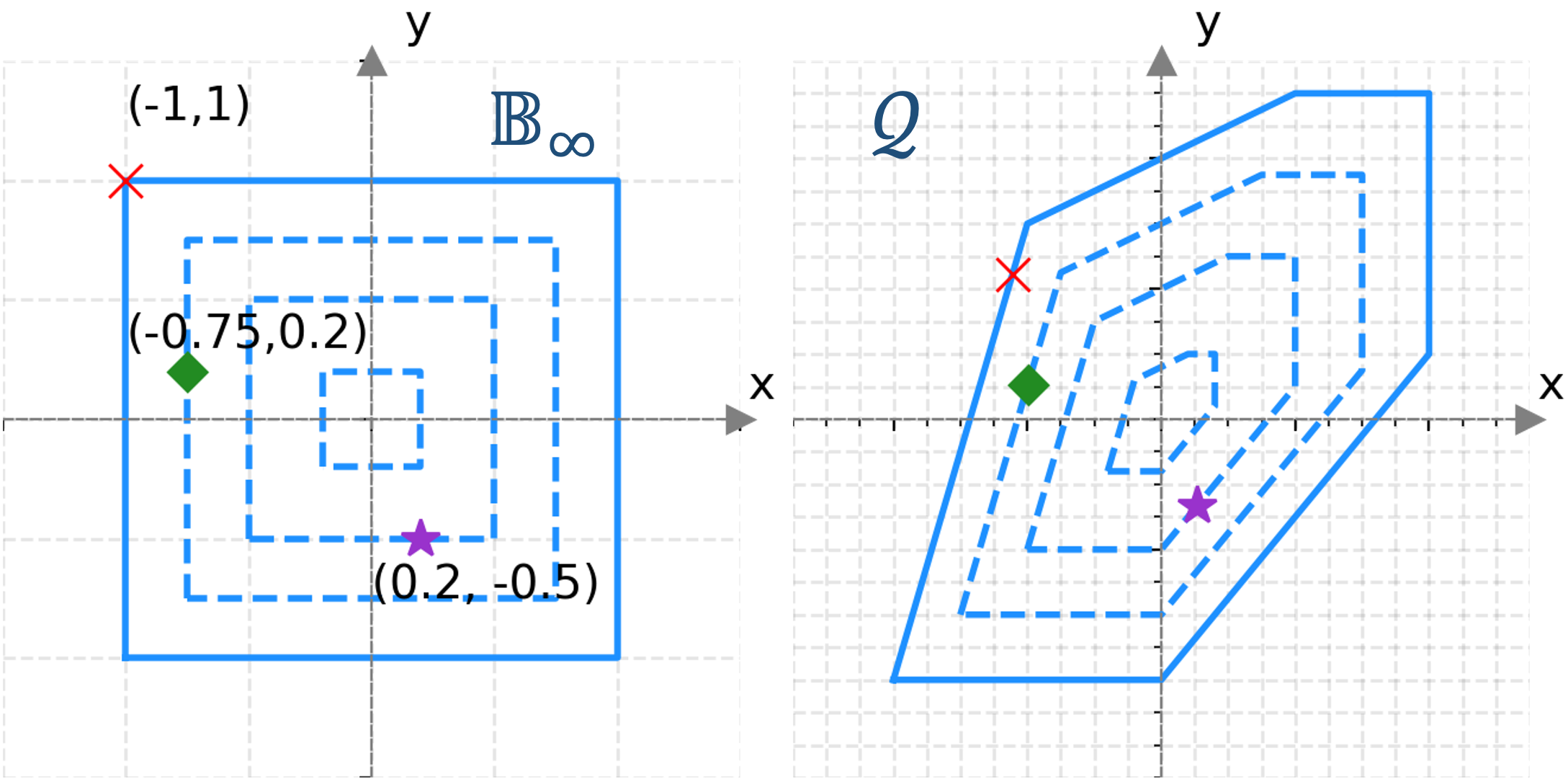}
\caption{ An illustrative example of the gauge map from $\mathbb{B}_{\infty}$ to a polyhedral C-set $\mathcal{Q}$.
The $1$, $\frac{3}{4}$, $\frac{1}{2}$ and $\frac{1}{5}$ level curves of each set are plotted in blue. For each point in $\mathbb{B}_{\infty}$, it is transformed to its image (marked using the same color) in $\mathcal{Q}$ with the same level curve.
\vspace{-0.5cm}}
\label{fig:gaugemap}
\end{figure}

\begin{definition}[Gauge function \cite{Blanchini07}]\label{def:gaugefuntion}
The gauge function associated with a C-set $\bP$ is a mapping $g_{\bP}: \R^{n} \longmapsto [0,+\infty]$, 
given by
\begin{equation}
g_{\bP}(\bz) = \min\{\lambda:\bz\in\lambda\bP,\lambda\geq 0,\bz\in\R^{n} 
\}.\nonumber
\end{equation}
\end{definition}
\begin{prop}\label{prop:gauge-polyhedral}
    If C-set $\bP$ is a polyhedral set of the form
    \begin{equation}
    \bP=\{\bz\in\R^{n}|\mathbf{A}\bz\leq\mathbf{b}, \mathbf{A}\in\R^{m\times n},\mathbf{b}\in\R^{n}\},\nonumber
    \end{equation}
    then the gauge function associated with it is
    \begin{equation}
    g_{\bP}(\bz)=\max_{i=1,\cdots,m}\big\{\frac{\mathbf{a}_i^T\bz}{b_i}\big\},\nonumber
    \end{equation}
    where $\mathbf{a}_i$ is the $i$-th row of $\mathbf{A}$ and $b_i$ is the $i$-th element of $\mathbf{b}$.
\end{prop}
The proof of Proposition \ref{prop:gauge-polyhedral} is provided in Appendix \ref{proof:gauge-polyhedral}. By using the gauge function defined in Definition \ref{def:gaugefuntion}, we can express the gauge map as follows.
\begin{definition}[gauge map\cite{Tabas2021ComputationallyES}]\label{def:gauge-map}
The gauge map between any two C-sets $\bP$ and $\bQ$ is a bijection $G: \bP\longmapsto\bQ$ given by 
\begin{equation}
    G(\bz|\bP, \bQ)=\frac{g_{\bP}(\bz)}{g_{{\bQ}}(\bz)}\bz.\nonumber
\end{equation}
\end{definition}
According to this definition, for each point $\bu$ within $\mathbb{B}_{\infty}$, it is possible to convert it into a feasible solution within $\Theta$, denoted as $\btheta_{u}$, by establishing a one-to-one correspondence (image) between $\mathbb{B}_{\infty}$ and $\Theta$ through the following gauge map:
\begin{equation}\label{eq:gauge-map}
    \btheta_{u} := G(\bu|\mathbb{B}_{\infty}, \Theta)=\frac{g_{\mathbb{B}_{\infty}}(\bu)}{g_{\Theta}(\bu)}\bu,
\end{equation}
where $g_{\mathbb{B}_{\infty}}(\bu)$ and $g_{\Theta}(\bu)$ can be calculated using Proposition \ref{prop:gauge-polyhedral}. Particularly, $g_{\mathbb{B}_{\infty}}(\bu)=\|\bu\|_{\infty}$. 
It is worth noting that the feasibility of the transformed point $\btheta_{u}$ is ensured by this definition of gauge map. To this end, we present the following theorem.
\begin{theorem}\label{thm:feasiility_ensuring}
    Given any point $\bu\in\mathbb{B}_{\infty}$, the transformed point $\btheta_{u}$ by the gauge map in \eqref{eq:gauge-map} lies within $\Theta$.
\end{theorem}
The proof of Theorem \ref{thm:feasiility_ensuring} can be found in Appendix \ref{appendix:prove_feasibility}.
Moreover, we provide an illustrative example in Fig.~\ref{fig:gaugemap}, visualizing the gauge transformation from $\mathbb{B}_{\infty}$ to a randomly generated polyhedral C-set.
\captionsetup[figure]{font=small,skip=2pt}

\begin{figure}[t]
\centering
\includegraphics[width=8cm, height=4.5cm]{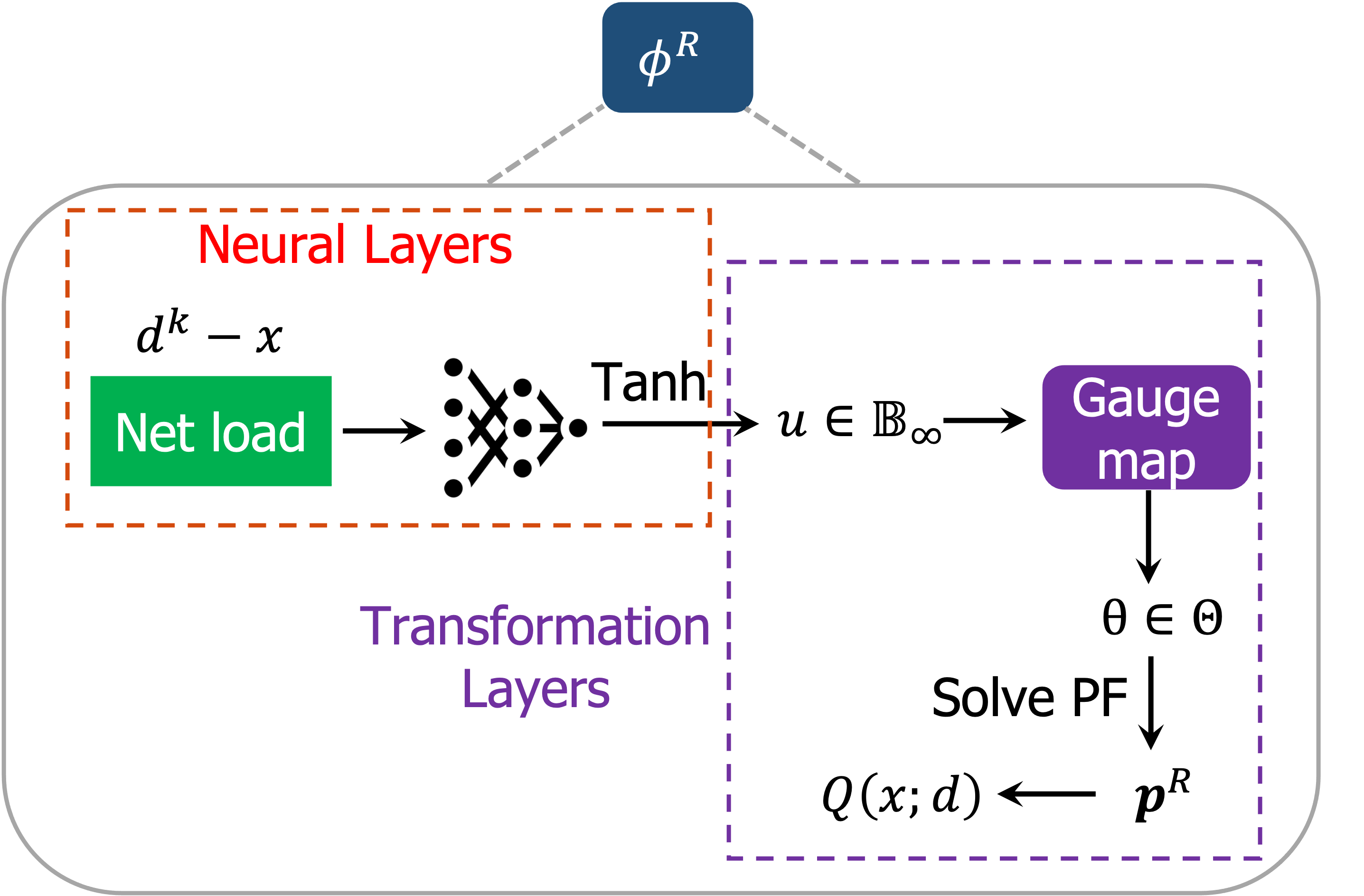}
\caption{
The hypercubic output from the neural layers is transformed to a feasible solution for $\btheta$ by the gauge map. Then the value of the objective function can be easily computed.
\vspace{-0.5cm}
}
\label{fig:archphiR}
\end{figure}

Once a feasible solution of $\btheta$ is obtained, the values for $\bp^{R}$ and the output of $\phi^{R}$, i.e., the objective value of the deterministic DCOPF in \eqref{generic-s2-k}, can be be easily computed. 
We summarize the network design of $\phi^{R}$ in Fig.~\ref{fig:archphiR}.

Lastly, we discuss the differentiability properties of the function in \eqref{eq:gauge-map} since training the network architecture in Fig.~\ref{fig:architecture_train}  requires a backward pass that can calculate the gradients in \eqref{eq:sgd}. This is a nuanced point since both \eqref{eq:gauge-map} and the layers used in neural networks are not everywhere differentiable. 
Here, we show that the non-differentiability introduced by the gauge map is no more severe than the non-differentiability that is already present in the neural network activation functions, and the end-to-end policy is differentiable almost everywhere:
\begin{theorem}\label{thm:subdifferential}
Let $\bP$ and $\bQ$ be polyhedral C-sets. Standard automatic differentiation procedures, when applied to the gauge map $G(\cdot \mid \bP, \bQ)$, will return the gradient of $G(\cdot \mid \bP, \bQ)$ for almost all $\textbf{z} \in \bP$. 
\end{theorem}
\begin{proof}
    The set $\bP$ can be partitioned such that the gauge map is a different analytic function on each region of the partition (excluding the origin). By setting $G(0 \mid \bP,\bQ) := 0$, we obtain a function for which
    standard automatic differentiation procedures will compute the gradient of $G(\cdot \mid \bP,\bQ)$ at all $\textbf{z} \in \bP$ except possibly on a set of measure zero~\cite{Lee2020}. Details are in Appendix \ref{appendix:autodiff}.
\end{proof}
Theorem \ref{thm:subdifferential} shows that the gauge map is differentiable with respect to the output of the neural layers, and hence enables the computation of backpropagation gradients in \eqref{eq:sgd} and the training of the architecture in Fig.~\ref{fig:architecture_train}. 
In the next section, we validate the effectiveness of the proposed learning architecture on a modified IEEE 118-bus system.

\captionsetup[figure]{font=small,skip=2pt}


\section{Experimental results}\label{sec:results}
In this section, we provide the experimental results of using the proposed algorithm in Table \ref{tab:algo1} to solve two-stage DCOPF problems. { Particularly, we consider two application contexts, namely, the risk-limiting dispatch and reserve scheduling problems on two systems, the IEEE 118-bus system~\cite{Burkhsh2013} and a 2000-bus synthetic grid based on ERCOT~\cite{activsg2000}.} Our algorithm learns the first-stage solutions to the scenario-based problems in \eqref{saa-origModel}  and \eqref{saa-resModel}, respectively. We implement our learning algorithm in Google Colab~\cite{colab} using Pytorch and all codes and data of our experiments are available at \url{https://github.com/zhang-linnng/two-stage-dcopf-neural-solver.git}.

\textbf{Network architecture:}
Two convolutional neural networks are constructed, with one representing $\phi^{0}$ and the other representing $\phi^{R}$. Each network consists of a $4$-layer structure, including 2 convolutional layers followed by 2 fully connected layers.
A dropout layer with the rate of $0.5$ is used on each of the fully connected layer before the output. 
The network architectures are trained offline using Adam Optimizer~\cite{adam} and the default learning rate is adopted.
The size of the hidden layer is tuned for each application context and the details can be found in our public code repository.

\textbf{Data generation:}
There are two types of data in our algorithm.
The first type is the load forecasts. They are inputs to the learning algorithm and comprise the datasets on which we train and test the network architecture. In both application contexts, the training dataset consists of $50000$ load forecasts and testing dataset of $100$.~\footnote{{The reason for limiting our testing to a small dataset in these experiments is not because of our learning algorithm, rather it was due to the high computational cost of generating benchmark solutions using CVXPY. As we will see, obtaining $100$ benchmark solutions for the reserve scheduling problem in a $118$-bus network took approximately $5$ hours. By contrast, since our model directly optimizes the objective function, we do not need ground truth data for training, so we can have a large training set. }}
The second type of data is the load realizations that are used to solve the scenario-based problems (estimating the expected second-stage cost) or to evaluate the solution quality through ex-post \textit{out-of-sample} simulations~\cite{roald2023}. 
{
During training, after experimenting with various load realization sampling sizes, we found that using $20$ load realizations led to a more notable reduction in the training loss over a fixed time period. Therefore, we opt to sample $20$ load realizations independently for each batch to approximate the expected second-stage cost.
For testing, the first stage solutions obtained using different methods are evaluated on a shared set of load realizations. To ensure a precise evaluation of the solution quality, we use $500$ load realizations for out-of-sample simulations.
}

Both types of data are generated using the Gaussian distribution but with different choices of the mean and standard deviation. When generating load forecasts, we use the base load of the system as the mean and set the standard deviation to be $10\%$ of it.
{To solve the two-stage problem for a given load forecast, we sample load realizations from a Gaussian distribution centered around this forecast, with a standard deviation of $5\%$ of this forecasted load value.}

\textbf{Baseline solvers:}
In both application contexts, we call the solver from CVXPY ~\cite{diamond2016cvxpy} to solve the scenario-based problems in \eqref{saa-origModel}  and \eqref{saa-resModel} on the testing dataset to establish a benchmark.
We also compare the solutions produced by our method to that by using the affine policy method, which is a widely applied approximation policy to make the two-stage stochastic programs tractable~\cite{bental2004affinepolicy}. {Specifically, the affine policy approximates the recourse decision $\mathbf{p}^{R}$ by representing it as an affine function of the uncertain parameter. This allows it to solve a simplified version of the original problem and save solving time. Further information on how to implement the affine policies for each specific application can be found in Appendix~\ref{appendix:affine-policy}.}

\textbf{Evaluation procedure:}
{For each testing instance, we apply various methods to produce their respective first-stage solutions.
To evaluate the quality of a specific first-stage solution, we compute the total cost arising from it by aggregating both the first-stage cost and the corresponding cost-to-go. The cost-to-go associated with a given first-stage solution is derived by solving the second-stage problem defined in \eqref{generic-s2-k} over a common set of $500$ load realizations. We then compare the performance of different methods based on their resulting average total costs and solving times on the testing dataset.
}


\subsection{Application \Romannum{1}: Risk-Limiting Dispatch}
{The results of using different methods to solve the risk-limiting dispatch problem in \eqref{saa-origModel} on the 118-bus system and the 2000-bus system are provided in Table~\ref{tab:applicationI-results} and Table~\ref{tab:applicationI-results-2000}, respectively.}
The average total costs of different methods are compared against the cost values produced by CVXPY.
{In both cases, our learning method is faster than applying CVXPY solver by 6 orders of magnitude while the difference in average total cost is less than $2\%$. In comparison, using the affine policy reduces the average running time by half, however, it also performs $20\%$ to $100\%$ worse. This is because the affine policy has bad generalization when applied to never-seen instances of load forecasts, and tend to resort to load shedding in the second stage, leading to high costs.}

\captionsetup[table]{font=small,skip=2pt}
\begin{table}[ht]
\normalsize
\centering
\begin{tabular}{lcc}
\hline
\multicolumn{3}{c}{\textbf{Risk-limiting dispatch on 118-bus system}}\\
\hline
\multirow{2}{*}{Methods}& {Increase in Total cost} & {Solving Time}\\
& (average, \%) & (average, seconds) \\
\hline
CVXPY & - & 24 \\
\textbf{Proposed} & \textbf{0.77} & $\mathbf{1\cdot 10^{-7}}$\\
Affine policy & 99.4 &  1.2 \\
\hline
\end{tabular}
\caption{
Performances of our method and the affine policy method to solve the risk-limiting dispatch problem in \eqref{saa-origModel} on the $118$-bus system.
The resulting total costs and solving times are averaged out over $100$ test instances and compared against the benchmark solutions obtained from CVXPY.
\vspace{-0.5cm}} 
\label{tab:applicationI-results}
\end{table}

\captionsetup[table]{font=small,skip=2pt}
\begin{table}[ht]
\normalsize
\centering
{
\begin{tabular}{lcc}
\hline
\multicolumn{3}{c}{\textbf{Risk-limiting dispatch on 2000-bus system}}\\
\hline
\multirow{2}{*}{Methods}& {Increase in Total cost} & {Solving Time}\\
& (average, \%) & (average, seconds) \\
\hline
CVXPY & - & 71 \\
\textbf{Proposed} & \textbf{1.5} & $\mathbf{1.2\cdot 10^{-4}}$\\
Affine policy & 19.6 &  3.6 \\
\hline
\end{tabular}}
\caption{
    {Compare the performance of using our method and the affine policy method to solve risk-limiting dispatch on the $2000$-bus system.
    The resulting total costs and solving times are averaged out over $1000$ test instances and compared against the benchmark solutions obtained from CVXPY.}
     } 
    \label{tab:applicationI-results-2000}
    \end{table}

\subsection{Application \Romannum{2}: Reserve Scheduling}
We summarize the results of using different methods to solve the reserve scheduling problem in \eqref{saa-resModel} on the 118-bus system in Table \ref{tab:applicationII-results}. 
All reported total costs are normalized in reference to the cost values obtained from CVXPY. Compared to the risk-limiting dispatch problem, the reserve scheduling problem has more decision variables and constraints, which make it more challenging to solve. It takes minutes to solve a single instance in CVXPY. By using an affine policy to approximate the recourse dispatch decision, the average running time per instance can be reduced by an order of magnitude, but the average total cost also increases by an order of magnitude due to poor generalization. 
In contrast, our learning method not only learns to provide good solution quality (within $10\%$ of the benchmark produced by CVXPY solver) but is also able to speed up the computation by 4 orders of magnitude.

\captionsetup[table]{font=small,skip=2pt}
\begin{table}[ht]
\normalsize
\centering
\begin{tabular}{lcc}
\hline
\multicolumn{3}{c}{\textbf{Reserve scheduling on 118-bus system}}\\
\hline
\multirow{2}{*}{Methods}& {Total cost} & {Solving Time}\\
& (average, \%) & (average, minutes) \\
\hline
CVXPY & 100 & 3.210 \\
\textbf{Proposed} & \textbf{110} & $\mathbf{10^{-4}}$ \\
Affine policy & 1813 &  0.343 \\
\hline
\end{tabular}
\caption{
Compare the performance of using our method and the affine policy method to solve the reserve scheduling problem in \eqref{saa-resModel} on the $118$-bus system.
The resulting total costs and solving times are averaged out over $100$ test instances and compared against the benchmark solutions obtained from CVXPY.
} 
\label{tab:applicationII-results}
\end{table}

Next, we compare our method with the widely used K-means scenario reduction method. Specifically, we reduce a collection of scenarios (load realizations) into two reduced sets: one with $5$ scenarios and the other with $2$ scenarios. To generate load realizations, we use a Gaussian distribution centered around the input forecast, with standard deviation of $50\%$ of this forecast.

We assess the performance of our method and the K-means scenario reduction approach across different experimental configurations by comparing their average total costs and solving times across $100$ test instances against the benchmark solutions obtained using CVXPY. The results are detailed in Table~\ref{tab:sr-results}. Particularly, all cost values are normalized relative to CVXPY's solutions, and we report the average increase in total cost within the table.

As we can see, our method achieves the smallest increase in total cost while requiring the least amount of time across both datasets with small and large load variations. Notably, in order to achieve a solving time comparable to ours, the scenario reduction method needs to shrink the scenario set by a factor of 10, i.e, to 2 scenarios. However, this considerable reduction significantly diminishes its performance when contrasted with both our method and the case using a reduced set of 5 scenarios. On the other hand, although using a factor of 10, corresponding to 5 scenarios, gives improved performance for the K-means scenario reduction method compared to the use of 2 scenarios, the resulting average total cost remains higher than our method's, and the solving speed is considerably slower, being $100$ times slower than ours.

     \captionsetup[table]{font=small,skip=2pt}
    \begin{table}[ht]
    \normalsize
    \centering
    {
    \begin{tabular}{lcc}
    \hline
    \multicolumn{3}{c}{\textbf{Comparions with Scenario Reduction }}\\
    \hline 
    \hline
    \multirow{2}{*}{Methods}& {Increase in Total cost} & {Solving Time} \\
    & (average, \%) & (average, seconds) \\
    \hline
    CVXPY &  - & 29 \\
    \textbf{Proposed} &   \textbf{2.8}& $\mathbf{6 \cdot 10^{-4}}$\\
     2 scenarios  &   6.6 &  $3\cdot{10^{-2}}$\\
     5 scenarios  &   4.4&  ${6 \cdot 10^{-2}}$\\
    \hline
    \end{tabular}}
    \caption{ 
     {Compare the performance of using our method and the K-means scenario reduction method for the 118-bus system. The resulting average total costs and solving times across $100$ test instances are compared against the benchmark solutions obtained from CVXPY.}
     } 
    \label{tab:sr-results}
    \end{table}

\section{Conclusions and Future work}\label{sec:conclusion}
This paper presents a learning algorithm to solve two-stage DCOPF problems efficiently. The algorithm {uses} two neural networks, one for each stage, to make the dispatch decisions. 
{The gauge map technique is built into the network architecture design 
so that the constraints in two-stage DCOPF problems can be guaranteed to be satisfied. This allows the proposed model to be trained in an unsupervised way without the need for generating ground truth data. }
The numerical results on the IEEE 118-bus and {2000-bus systems}  validate the effectiveness of the proposed algorithm, showing that it can speed up computation by orders of magnitude compared to the commercial solver while still learning high-quality solutions. A direction of future work is to generalize our learning algorithm to solve non-convex programs, for example, using the AC optimal power flow model.

\appendices
\section{Expressions of $\bB$,  $\mathbf{F}$ and $\mathbf{E}$}\label{appendix:BF}
Suppose we use $\mathcal{E}$ to denote the set of all lines in the power system and $(i,j)$ the line connecting bus-$i$ and bus-$j$. Without loss of generality, we can assume the line $(i,j)$ is the $m$-th out of all lines. 
Let $b_{ij}$ be the susceptance for the line $(i,j)$, then the flow on line $(i,j)$ is $f_{ij} = b_{i,j}(\theta_i-\theta_j)$. The nodal power injection by bus-$i$ is $p_{i}=\sum_{k:(i,k)\in\mathcal{E}}f_{ik}=\sum_{k:(i,k)\in\mathcal{E}}b_{i,j}(\theta_i-\theta_j)$.
As a result, the matrix $\mathbf{B}$ that transforms the phase angle $\btheta\in\R^{N-1}$ into the nodal power injections at all buses can be expressed as
\begin{equation}\nonumber
   \forall i,j: B_{ij}=\left\{
\begin{array}{ll}
      -b_{ij},\text{ if } (i,j)\in\mathcal{E} \text{ and } i\neq j  \\
      \sum_{k:(k,j)\in\mathcal{E}}b_{kj},\text{ if } (i,j)\in\mathcal{E} \text{ and } i=j \\
      0, \text{ otherwise}. 
\end{array} 
\right. 
\end{equation}
The matrix $\mathbf{F}$ that maps $\btheta$ to flows on all lines is given by
\begin{equation}\nonumber
\begin{array}{l}
  F_{mi}=b_{ij}, F_{mj}=-b_{ij},\\
  \forall m\in\{1,\cdots,M\}, i,j\in\{1,\cdots,N\} \text{ and } i\neq j.
\end{array} 
\end{equation}
{The incidence matrix $\mathbf{E}$ has a row for each node and a column for each line.
To construct $\mathbf{E}$, we can treat the power network as a directed graph, where each line has a starting node (source) and an ending node (target),
For instance, consider a line denoted by $(i,j)$ as the $m$-th among all lines, with $i>j$. In this setup, we assume that the node with a smaller index, i.e., node $i$, serves as the source node, while node $j$ is designated as the target node. Accordingly, the entry $\mathbf{E}_{[[m,i]]}$ in $\mathbf{E}$ is assigned a value of $1$, and the entry $\mathbf{E}_{[[m,j]]}$ is assigned $-1$.}

\section{SGD Updating Rules}\label{appendix:sgd}
{
The stochastic gradients of the loss function with respect to $\bw^{0}$ and $\bw^{R}$ at a randomly chosen data point $\widebar{\bd}^{i}$ are calculated using the chain rule in the backward pass, which can be expressed as follows
\begin{subequations}\label{eq:sgd}
\begin{align}
&{\frac{\partial L^{i}(\bw^{0},\bw^{R})}{\partial \bw^{0}}}=~
{\widetilde{c}}^{\prime}\frac{\partial \phi^{0}(\widebar{\bd}^{i};\bw^{0})}{\partial \bw^{0}} + \nonumber\\
&\frac{1}{K}\sum_{k=1}^{K} \frac{ \partial \phi^{R}(\tilde{\bdelta}_d^{ik};\bw^{R}) }{\partial \tilde{\bdelta}_d^{ik}} \frac{\partial \tilde{\bdelta}^{ik}}{\partial \phi^{0}(\widebar{\bd}^{i};\bw^{0})} \frac{\partial\phi^{0}(\widebar{\bd}^{i};\bw^{0})}{\partial \bw^{0}}\label{eq:sgd-w0}\\
&{\frac{\partial L^{i}(\bw^{0},\bw^{R})}{\partial \bw^{R}}}=\frac{1}{K}\sum_{k=1}^{K} \frac{ \partial \phi^{R}(\tilde{\bdelta}_d^{ik};\bw^{R}) }{\partial \bw^{R}}.\label{eq:sgd-wR}
\end{align}
\end{subequations} 
where 
$\widetilde{c}^{\prime}$ is the derivative of $\widetilde{c}(\cdot)$.
At each iteration $t$, SGD repeats the following updates on $\bw^{0}$ and $\bw^{R}$ until a certain stopping criterion is reached:
\begin{subequations}\label{eq:sgd-update}
    \begin{align}
{\bw^{0}}^{(t+1)}&\longleftarrow{\bw^{0}}^{(t)}-\rho {\frac{\partial L^{i}(\bw^{0},\bw^{R})}{\partial \bw^{0}}}\Big{|}_{{\bw^{0}}^{(t)}}\\
{\bw^{R}}^{(t+1)}&\longleftarrow{\bw^{R}}^{(t)}-\rho {\frac{\partial L^{i}(\bw^{0},\bw^{R})}{\partial \bw^{R}}}\Big{|}_{{\bw^{R}}^{(t)}},
    \end{align}
\end{subequations}
where $\rho$ denotes the step size. Note that all the backward pass gradients given by \eqref{eq:sgd} can be computed using the automatic differentiation engine in machine learning libraries, such as \textit{autograd} in Pytorch~\cite{paszke2017automatic, NEURIPS2019PyTorch}, and the SGD updating rules in \eqref{eq:sgd-update} can also be implemented therein. 
}

\section{Proof of Theorem \ref{thm:Cset} }\label{proof:Cset}
To show that the polyhedron given by \eqref{def:Theta} is bounded, we use the definition of a bounded polyhedra: \textit{a polyhedra is bounded if $\exists K>0$ such that $\|\btheta\|\leq K$, for all $\btheta\in\Theta$}. 

From Appendix \ref{appendix:BF}, we know that the flow on line $(i,j)$ can be expressed as $f_{ij} = b_{i,j}(\theta_i-\theta_j)$; therefore, we can rewrite the polyhedra in \eqref{def:Theta} as
\begin{equation}\nonumber
    -f_{i,j}^{\max}\leq b_{i,j}(\theta_i-\theta_j)\leq f_{i,j}^{\max}, \forall (i,j)\in\mathcal{E},
\end{equation}
which are equivalent to
\begin{equation}\label{constr:flow_limit}
    \frac{-f_{i,j}^{\max}}{b_{i,j}}\leq \theta_i-\theta_j\leq \frac{f_{i,j}^{\max}}{b_{i,j}}, \forall (i,j)\in\mathcal{E},
\end{equation}
that is, both $\theta_i$ and $\theta_j$ must be bounded, otherwise, \eqref{constr:flow_limit} would be violated. Since every bus in the system must be connected to at least one other bus, \eqref{constr:flow_limit} implies that $\exists K_i\in\R_{+}$ such that $|\theta_i|\leq K_i, \forall i\in\{1,\cdots,N\}$, therefore, we can choose $K=\max_i\{K_i\}$ and then we have $\|\btheta\|\leq K$. By definition, the polyhedra given by \eqref{def:Theta} is bounded.

\section{Proof of Proposition \ref{prop:gauge-polyhedral}}\label{proof:gauge-polyhedral}
By Definition \ref{def:gaugefuntion}, we can express the gauge function associated with the polyhedral set $\bP=\{\bz\in\R^{n}|\mathbf{A}\bz\leq\mathbf{b}, \mathbf{A}\in\R^{m\times n},\mathbf{b}\in\R^{n}\}$ as the optimization problem $g_{\bP}(\bz)=\min\{\lambda:\mathbf{A}\bz\leq\lambda\mathbf{b},\lambda\geq 0,\bz\in\R^{n}\}$, which is equivalent to finding a non-negative value of $\lambda$ such that $\mathbf{a}_{i}^{T}\bz\leq \lambda b_i, \forall i\in\{1,\cdots,m\}$, that is, $\lambda\geq \frac{\mathbf{a}_{i}^{T}\bz}{b_i}, \forall i\in\{1,\cdots,m\}$. Therefore, the optimal value of $\lambda$, namely, the value of the gauge function $g_{\bP}(\bz)$, can be given by $\max_{i=1,\cdots,m}\big\{\frac{\mathbf{a}_i^T\bz}{b_i}\big\}$.

\section{Proof of Theorem \ref{thm:feasiility_ensuring}}\label{appendix:prove_feasibility}
{Since 
    $g_{\Theta}(\bu)=\max_{i=1,\cdots,m}\big\{\frac{\mathbf{a}_i^T\bu}{b_i}\big\}$, it follows that $\frac{\mathbf{a}_k^T\bu}{b_k}\leq g_{\Theta}(\bu)$, which is equivalent to $\frac{1}{g_{\Theta}(\bu)}\mathbf{a}_k^T\bu \leq b_k$, for any index $k$.  Given that $\bu$ lies within the unit hypercube $\mathbb{B}_{\infty}\triangleq\{\bz\in\R^{n}|-1\leq z_i\leq1,\forall i\}$, it follows that  $\|\bu\|_{\infty}\leq 1$. Bringing these together, we have $\frac{\|\bu\|_{\infty}}{g_{\Theta}(\bu)}\mathbf{a}_k^T\bu \leq b_k$, for any $k$.
    This series of inequalities can be compactly expressed as $\frac{\|\bu\|_{\infty}}{g_{\Theta}(\bu)}{\mathbf{A}}\bu \leq \mathbf{b}$. 
    By using \eqref{eq:gauge-map}, we can then conclude that ${\mathbf{A}}\btheta^{u}\leq \mathbf{b}$, which indicates that $\btheta^{u}$ is a feasible point lying within the set $\Theta$.
    }

\section{Proof of Theorem \ref{thm:subdifferential}} \label{appendix:autodiff}

Let $\bP = \{\textbf{z} \in \R^n \mid \textbf{A}\textbf{z} \leq \textbf{b}\}$ and $\bQ = \{\textbf{z} \in \R^n \mid \textbf{Cz} \leq \textbf{d}\}$ with $\textbf{A} \in \R^{m \times n}, \textbf{b} \in \R^{m}_{++}, \textbf{C} \in \R^{k \times n}$, and $\textbf{d} \in \R^k_{++}$. Let $\mathcal{A}_{ij}$ be the polytope described as $\{\textbf{z} \in \bP \mid i \in \argmax_{i=1,\cdots,m} \frac{\textbf{a}_i^T\textbf{z}}{b_i}, j \in \argmax_{j=1,\cdots,k} \frac{\textbf{c}_j^T\textbf{z}}{d_j}\}$. The set $\{\mathcal{A}_{ij} \mid i \in 1,\cdots,m,\ j \in 1,\cdots,k\}$ forms a polyhedral partition of $\bP$, and the gauge map is an analytic function on the interior of each $\mathcal{A}_{ij}$ except when $\textbf{c}_j^T\textbf{z} = 0$ or $\textbf{z} = 0$. Specifically, the gauge map on the interior of $\mathcal{A}_{ij} \subseteq \bP$ can be written as $G(\textbf{z}\mid \bP,\bQ) = \frac{\textbf{a}_i^T\textbf{z}/b_i}{\textbf{c}_j^T\textbf{z}/d_j}\textbf{z}$. 

For any $j \in 1,\cdots,k$, $\textbf{c}_j^T\textbf{z} = 0$ if and only if $\textbf{z} = 0$: since $Q$ forms a full-dimensional and bounded polytope, $\textbf{C}$ must be full-rank and tall ($k>n$). Thus, $\textbf{Cz} = 0$ if and only if $\textbf{z} = 0$. 

We can now justify the choice $G(0 \mid \bP,\bQ) := 0$ as follows. Let $\textbf{z} = \alpha \textbf{h}$ for some $\alpha > 0$ and $\textbf{h} \in \R^n \backslash \{0\}.$ There exist some $(i,j)$ and sufficiently small $\varepsilon > 0$ such that $\textbf{z} \in \mathcal{A}_{ij} \ \forall\ \alpha \in (0,\varepsilon).$ The limit of $\frac{\textbf{a}_i^T\textbf{z}/b_i}{\textbf{c}_j^T\textbf{z}/d_j}\textbf{z}$ as $\alpha \rightarrow 0$ is equal to $0 \in \R^n$.  

By the above analysis, the gauge map is \textit{piecewise analytic under analytic partition} (PAP) on $\bP$ which implies desirable properties for automatic differentiation \cite{Lee2020}. Specifically, PAP functions can be composed with one another (they obey a chain rule), they are differentiable almost everywhere (except possibly on a set of measure zero), and standard automatic differentiation tools will compute the derivatives at all points where the function is differentiable.

\section{Formulation of Affine Policy}\label{appendix:affine-policy}
{Given a first stage dispatch $\bp^{0}$, we
adopt the following affine policy to approximate the recourse dispatch ${\bp^{R}}({\bomega}^{k})$ by representing it as 
an affine function of the load realization $\bd({\bomega}^{k})$:}
\begin{equation}\label{eq:affine-policy}
{\bp^{R}}({\bomega}^{k}) = \bxi(\mathbf{1}^{T}\bd({\bomega}^{k})-\mathbf{1}^{T}\bp^{0}).
\end{equation}
Here, $\bxi=[\xi_1,\cdots,\xi_N]^{T}\in\R^{N}$ denotes the participation factor vector, which represents the contribution of each generator to balancing the discrepancy between the actual load and the first-stage dispatch. These participation factors are subject to the following constraints:
\begin{equation}\label{constr:dist_factor}
\xi_i \geq 0, \forall i\in\mathcal{G}, \; 
\xi_i=0, \forall i \in\mathcal{N}/\mathcal{G}, \;
\sum_{i\in\mathcal{G}}\xi_i=1, 
\end{equation}
where the set $\mathcal{G}$ represents all buses hosting generators, and the set $\mathcal{N}/\mathcal{G}$ includes all buses except for those connected to generators. Note that the values of $\bxi$ are determined during the first stage and are not influenced by the particular load realization in the second stage.

{
To implement the affine policy, we start by substituting the variable ${\bp^{R}}({\bomega}^{k})$ with the affine function \eqref{eq:affine-policy} in the original problem formulations \eqref{saa-origModel} and \eqref{saa-resModel}. This enables us to solve the following approximation problems to determine the values of $\bxi$:}

\textbf{Approximation problem of risk-limiting dispatch:}

\begin{subequations} \label{rld-affinepolicy}
\begin{align} 
\min_{
\substack{\bp^{0},\bxi\\ \{{\btheta}({\bomega}^{k})\}_{k=1}^{K}}
}~& \balpha^T\bp^{0} + \frac{1}{K}\sum_{k=1}^{K} \bbeta^T[{\bxi(\mathbf{1}^{T}\bd({\bomega}^{k})-\mathbf{1}^{T}\bp^{0})}]^{+} \nonumber\\
\textrm{s.t.~} & \eqref{3a}-\eqref{3d}, \eqref{eq:affine-policy}, \eqref{constr:dist_factor}\nonumber.
\end{align}
\end{subequations}

\textbf{Approximation problem of reserve scheduling:}
\begin{subequations} 
\begin{align}
\min_{
\substack{\bp^{0},\widehat{\br},\check{\br},\\ \bxi,\{{\btheta}({\bomega}^{k})\}_{k=1}^{K}}
}~&\balpha^T\bp^{0} + \bmu^T(\widehat{\br}+\check{\br})~+\nonumber\\
\quad\frac{1}{K}\sum_{k=1}^{K} \Bigg(
{\bgamma^{\res}}^{T}
& \Big( [\bxi(\mathbf{1}^{T}\bd({\bomega}^{k})-\mathbf{1}^{T}\bp^{0})-\widehat{\br}]^{+}- \nonumber\\
& [\bxi(\mathbf{1}^{T}\bd({\bomega}^{k})-\mathbf{1}^{T}\bp^{0})+\check{\br}]^{-} \Big)
\Bigg)\nonumber\\
\textrm{s.t.~} & \eqref{6a}-\eqref{6g}, \eqref{eq:affine-policy}, \eqref{constr:dist_factor}\nonumber.
\end{align}
\end{subequations}
Once determined, $\bxi$ is treated as fixed and  we solve the following simplified linear programs to obtain the first-stage decisions for a set of anticipated load realizations:

\textbf{Implement affine policy for risk-limiting dispatch:}
\begin{align} \label{rld-affinepolicy2}
\min_{
\substack{\bp^{0},\\ \{{\btheta}({\bomega}^{k})\}_{k=1}^{K}}
}~& \balpha^T\bp^{0} + \frac{1}{K}\sum_{k=1}^{K} \bbeta^T[{\bxi(\mathbf{1}^{T}\bd({\bomega}^{k})-\mathbf{1}^{T}\bp^{0})}]^{+} \\
\textrm{s.t.~} & \eqref{3a}-\eqref{3d}, \eqref{eq:affine-policy}\nonumber.
\end{align}

\textbf{Implement affine policy for reserve scheduling:}
\begin{align}\label{reserve-affinepolicy2}
\min_{
\substack{\bp^{0},\widehat{\br},\check{\br},\\ \{{\btheta}({\bomega}^{k})\}_{k=1}^{K}}
}~&\balpha^T\bp^{0} + \bmu^T(\widehat{\br}+\check{\br})~+\nonumber\\
\quad\frac{1}{K}\sum_{k=1}^{K} \Bigg(
{\bgamma^{\res}}^{T}
& \Big( [\bxi(\mathbf{1}^{T}\bd({\bomega}^{k})-\mathbf{1}^{T}\bp^{0})-\widehat{\br}]^{+}- \nonumber\\
& [\bxi(\mathbf{1}^{T}\bd({\bomega}^{k})-\mathbf{1}^{T}\bp^{0})+\check{\br}]^{-} \Big)
\Bigg)\\
\textrm{s.t.~} & \eqref{6a}-\eqref{6g}, \eqref{eq:affine-policy}\nonumber.
\end{align}
In contrast to the original problems presented in \eqref{saa-origModel} and \eqref{saa-resModel}, the linear programs in \eqref{rld-affinepolicy2} and \eqref{reserve-affinepolicy2} reduce numbers of both decision variables and constraints. As a result, these simplified linear programs can be solved more quickly.







\ifCLASSOPTIONcaptionsoff
  \newpage
\fi



%

\bibliography{bibtex/bib/mybib}{}
\bibliographystyle{IEEEtran}




%








\end{document}